\documentclass[a4paper, 10 pt, conference]{ieeeconf}

\usepackage{graphicx}
\usepackage{balance}
\usepackage{comment}
\usepackage{cite}
\usepackage{amssymb}
\usepackage[tight,footnotesize]{subfigure}
\usepackage[active]{srcltx}
\usepackage{amsmath}
\usepackage{mathtools}

\graphicspath{{./figs/}}
\renewcommand{\baselinestretch}{0.95}
\usepackage{eurosym}
\usepackage[plain]{algorithm}
\usepackage{algorithmic}
\usepackage{multicol}
\usepackage{dsfont}
\usepackage{amsfonts}

\usepackage{cases}
\usepackage{xcolor}
\DeclareMathOperator{\tr}{tr}

\IEEEoverridecommandlockouts
\begin{document}

\title{Joint Estimation and Control for Multi-Target Passive Monitoring \\ with an Autonomous UAV Agent}

\author{Savvas~Papaioannou,~Christos~Laoudias,~Panayiotis~Kolios,\\~Theocharis~Theocharides and~Christos~G.~Panayiotou
\thanks{This work is co-funded by the Ministry of Defence of the Republic of Cyprus, the European Union’s Horizon 2020 research and innovation programme under grant agreement No. 739551 (KIOS CoE), and the Government of the Republic of Cyprus through the Cyprus Deputy Ministry of Research, Innovation and Digital Policy. 
The authors are with the KIOS Research and Innovation Centre of Excellence (KIOS CoE) and the Department of Electrical and Computer Engineering, University of Cyprus, Nicosia, 1678, Cyprus. {\tt\small \{papaioannou.savvas, laoudias, pkolios, ttheocharides, christosp\}@ucy.ac.cy}}
}

\maketitle

\begin{abstract}
This work considers the problem of passively monitoring multiple moving targets with a single unmanned aerial vehicle (UAV) agent equipped with a direction-finding radar. This is in general a challenging problem due to the unobservability of the target states, and the highly non-linear measurement process. In addition to these challenges, in this work we also consider: a) environments with multiple obstacles where the targets need to be tracked as they manoeuvre through the obstacles, and b) multiple false-alarm measurements caused by the cluttered environment. To address these challenges we first design a model predictive guidance controller which is used to plan hypothetical target trajectories over a rolling finite planning horizon. We then formulate a joint estimation and control problem where the trajectory of the UAV agent is optimized to achieve optimal multi-target monitoring.
\end{abstract}

\section{Introduction} \label{sec:Introduction}
Increased mobility, flexibility, and rapid deployment are highly desirable properties in many application domains. Nowadays, unmanned aerial vehicles (UAVs) have demonstrated their potential in a wide variety of applications including surveillance and security tasks \cite{Huang2020,SP3,SP4,SP5}, search-and-rescue missions \cite{Erdelj2017,SP1,SP2}, and situational awareness for first responders\cite{Moon2021}. To meet the requirements in the above-mentioned applications, the UAV agents often carry on-board various sensors such as direction-finders that passively scan the spectrum to detect and resolve the direction of a target transmission, and radars that provide the direction and/or distance by processing the reflections on objects by either purposefully transmitted signals (i.e., \textit{active} radar) or by ambient signals of opportunity (i.e., \textit{passive} radar).

Passive systems, e.g., based on radar technology \cite{griffiths2022introduction}, are less power demanding as no signal transmissions are required, thus extending the UAV agent's flight time. While this makes them preferable in many practical scenarios, such systems typically provide bearing-only measurements, i.e., the \textit{angle} between the target-agent line and a reference direction (e.g., magnetic north). In this case several challenges appear due to the highly nonlinear measurement process, and the unobservability of the target states, especially when a single UAV agent in considered.


Over the past years a plethora of approaches have been proposed in the literature for the problem of passive target monitoring/tracking. In this work we will focus mainly on the task of passive target monitoring with a single UAV agent which utilizes angle measurements (i.e., bearings), for estimating the target's state. A recent survey paper on this topic can be found in \cite{sindhu2019bearing}. Regarding the bearings-only passive target monitoring, the authors in \cite{la2008analysis} provide a thorough analysis of 3 different state-estimation approaches for tracking a single target with a single sensor. The authors in \cite{li2012particle} design a particle-filter based estimator that uses multiple radar measurements with glint noise in order to passively monitor a single moving target, and the work in \cite{yang2015robust} proposes a robust fuzzy extended-Kalman filter for monitoring a moving target. 

With respect to the agent/observer control aspect which appears in the passive target monitoring applications, the work in \cite{vander2014cautious} proposes a greedy algorithm for optimally choosing the measurement locations in order to localize a stationary target in the least amount of time. Similarly, in \cite{bayram2016gathering} the authors first use the geometric dilution of precision to characterise the uncertainty of passive target localization using angle measurements, and then they propose a measurement gathering strategy that jointly minimizes the target localization error of a stationary target, and the time spend in gathering the measurements. 
The problem of optimally controlling an autonomous agent/observer for accurate passive monitoring of a moving target is further investigated in \cite{arulampalam2004bearings,skoglar2009information}. Specifically, in \cite{arulampalam2004bearings} various particle-filter estimators are proposed based on the multiple model jump Markov system framework to tackle the various manoeuvres of the target, whereas in \cite{skoglar2009information} the observer control is posed as stochastic optimal control problem which aims at maximizing the tracking accuracy. Finally, the authors in \cite{oshman1999optimization} formulate the problem of observer control for bearings-only target localization as an optimal control problem which maximizes the determinant of the Fisher information matrix.

Complementary to the related works discussed above, in this paper we propose a multi-target passive-monitoring approach in which an autonomous UAV agent optimally decides its control inputs such that the combined uncertainty over the states of all targets is minimized. Contrary to existing solutions, this work investigates the problem in complex environments with obstacles, that need to be avoided by the targets, and which also cause multiple false-alarm measurements that need to be rejected by the estimator.

%

The rest of the paper is structured as follows. The system model is discussed in Section~\ref{sec:system_model}  and the proposed controller for planning the target trajectories is described in Section~\ref{sec:hypothesis_gen}. Section~\ref{sec:monitoring} presents the proposed joint estimation and control approach, and Section~\ref{sec:Evaluation} discusses the performance evaluation of the proposed approach. Finally, Section~\ref{sec:conclusion} provides concluding remarks.

\section{System Modelling} \label{sec:system_model}

\subsection{Target Dynamics} \label{ssec:target_dynamics}
In this work we assume that a known number of $M$ ground targets $\boldsymbol{x}^j, ~j \in [1,..,M]$ operate inside a bounded surveillance environment $\mathcal{E} \subset \mathbb{R}^3$ according to the following stochastic discrete-time dynamical model:
\begin{equation} \label{eq:target_dynamics}
    \boldsymbol{x}^j_{t} = A\boldsymbol{x}^j_{t-1} + B \boldsymbol{u}^j_{t-1} + \boldsymbol{\nu}_{t-1}, ~j \in [1,..,M]
\end{equation}

\noindent where $\boldsymbol{x}^j_t = [x^j_t(x),x^j_t(y),x^j_t(z), \dot{x}^j_t(x), \dot{x}^j_t(y),\dot{x}^j_t(z)]^\top \in \mathbb{R}^6$ denotes the state of the $j_\text{th}$ target at time $t$, which is composed of the target's position $(x^j_t(x),x^j_t(y),x^j_t(z))$, and velocity $(\dot{x}^j_t(x), \dot{x}^j_t(y),\dot{x}^j_t(z))$ components in 3D Cartesian coordinates. The control input $\boldsymbol{u}^j_t \in \mathbb{R}^3$ denotes the applied control force which allows the target to change its direction and speed, and the term $\boldsymbol{\nu}_t$ is the process noise which models the uncertainty on the target's state, and which is distributed according to a zero mean multi-variate Gaussian distribution with covariance matrix $Q$, i.e., $\boldsymbol{\nu}_t \sim \mathcal{N}(0,Q)$. Without loss of generality we assume that the process noise profile is the same for all targets. The matrices $A$ and $B$ are defined as:
\begin{equation}\label{eq:PhiGamma}
A = 
\begin{bmatrix}
    I_{3\times3} & \Delta t \cdot I_{3\times3}\\
    0_{3\times3} & (1-\varepsilon) \cdot I_{3\times3}
   \end{bmatrix},~
B = 
\begin{bmatrix}
    0_{3\times3} \\
     \frac{\Delta t}{m} \cdot I_{3\times3}
   \end{bmatrix},
\end{equation}

\noindent where $\Delta t$ is the sampling interval, $\varepsilon \in [0,1]$ models the effect of friction on the target's velocity, and $m$ is the target mass which for brevity we assume to be the same for all targets. Moreover, $I_{3 \times 3}$, and $0_{3 \times 3}$ denote the identity and zero matrices of size 3-by-3 respectively. 
Finally, it is assumed that during a reconnaissance phase, the approximate target initial location, and final destination have been acquired and made available to the UAV agent. Therefore, we assume that: a) target's $j$ initial state $\boldsymbol{x}^j_{0}$ is distributed according to $\boldsymbol{x}^j_{0} \sim \mathcal{N}(\boldsymbol{\mu}^j_0,\Sigma^j_0)$, and b) the target $j$ is moving towards a goal region on the ground denoted hereafter as $\mathcal{G}^j \subset \mathbb{R}^3$.


\subsection{Agent Dynamics} \label{ssec:agent_dynamics}
An autonomous UAV agent/observer, equipped with a passive direction-finding radar which is calibrated for a certain altitude $h$, is deployed inside the surveillance environment $\mathcal{E}$ with the purpose of monitoring the trajectories of the $M$ targets on the ground. The state of the UAV agent at time-step $t$ i.e., $\boldsymbol{s}_t = [s_t(x), s_t(y), s_t(z)]^\top \in \mathcal{E}$ which is composed of the agent's position in cartesian coordinates, evolves in time according to:
\begin{equation} \label{eq:controlVectors}
\begin{bmatrix}
s_t(x)\\
s_t(y)\\
s_t(z)
\end{bmatrix} 
\!\!=\!\!
\begin{bmatrix}
s_{t-1}(x)\\
s_{t-1}(y)\\
h
\end{bmatrix}
\!\!+\!\!
\begin{bmatrix}
\lambda\Delta_r \text{cos}(\kappa \Delta_\theta)\\
\lambda\Delta_r \text{sin}(\kappa \Delta_\theta)\\
0
\end{bmatrix}\!\!,\!\!
\begin{array}{l} 
\lambda \in  [0,..,N_r]\\ 
\kappa \in [1,..,N_\theta]
\end{array}\!\!\!, 
\end{equation}
where $\Delta_r$ is the radial step size, $\Delta_\theta=2\pi/N_\theta$, and the parameters $(N_r,N_\theta)$ specify the set $\mathcal{S}_t$ containing all possible states $\boldsymbol{s}_t \in \mathcal{S}_t$ which the agent can take at time-step $t$. Therefore, the set $\mathcal{S}_t$ is given by: $\mathcal{S}_t = \left\{\left(s_{t-1}(x)+\lambda\Delta_r \text{cos}(\kappa \Delta_\theta),s_{t-1}(y)+\lambda\Delta_r \text{sin}(\kappa \Delta_\theta),h\right)\right\}$, $\forall \lambda \in  [0,..,N_r], ~\forall \kappa \in [0,..,N_\theta]$.

\subsection{Agent Sensing Model} \label{ssec:agent_sensing_model}
As already mentioned the UAV agent is equipped with a passive radar (i.e., a direction-finder) which is used for  monitoring nearby ground targets operating inside its sensing range. Specifically, at each time-step $t$, the UAV agent receives a set of noisy angular measurements (i.e., bearings) from each target $j$, denoted as $\Phi^j_t = \{\phi^j_{t,1},..,\phi^j_{t,|\Phi^j_t|}\}, ~\phi^j_{t,i} \in (-\pi,\pi]$ rad, where the number of total received measurements, i.e., $|\Phi^j_t|$ ($|.|$ denotes the set cardinality), is random. In particular, it is assumed that due to various obstacles and clutter in the environment the UAV agent receives at each time-step $t$: a) with a Poisson rate $\Lambda$ multiple false-alarm measurements (denoted as $\tilde{\phi}^j_{t,i} \in \Phi^j_t$) which are distributed over the measurement space according to the probability distribution $p_{\tilde{\phi}}(\tilde{\phi}^j_{t,i})$, and b) a single bearing measurement $\hat{\phi}^j_{t} \in \Phi^j_t$ from the target $j$ with probability $p_D$. The target generated measurement $\hat{\phi}^j_{t}$ is related to the target and agent states according to the measurement model $\hat{\phi}^j_{t} = \ell(\boldsymbol{x}^j_t,\boldsymbol{s}_t) + w_t$, where:
\begin{equation} \label{eq:measurement_model}
	\ell(\boldsymbol{x}^j_t,\boldsymbol{s}_t) = \tan^{-1}\left(\frac{x^j_t(x)-s_t(x)}{x^j_t(y) - s_t(y)}\right),
\end{equation}
\noindent and $w_t$ is a Gaussian random variable which models the measurement noise, and which is distributed according to $w_t \sim \mathcal{N}(0,\sigma^2_\phi)$. Without loss of generality we assume that the same target detection probability, false-alarm rate, and the measurement noise applies for all targets, since all targets are sensed by the same radar equipment. In addition, we assume in this work that the targets are sensed by the UAV agent through different communication channels.

\subsection{Obstacle Model}\label{ssec:obstacle_model}
We consider the existence of multiple convex obstacles $\xi_n \in \Xi, ~ n \in [1,..,|\Xi|]$ inside the surveillance area $\mathcal{E}$, which are represented in this work as cuboids of arbitrary sizes. In particular, a regular cuboid $\xi$ is a box-shaped object with six rectangular faces, and 8 right angles; therefore, a point $\boldsymbol{p}=[p(x),p(y),p(z)]^\top \in \mathbb{R}^3$ that resides inside the convex-hull of cuboid $\xi_n$ must satisfy the following 6 linear inequalities:
\begin{align}
    a^n_{1}(x)p(x) + a^n_{1}(y)p(y) & + a^n_{1}(z)p(z)  \le b^n_1, \notag \\
    a^n_{2}(x)p(x) + a^n_{2}(y)p(y) & + a^n_{2}(z)p(z)  \le b^n_2, \notag \\
    \vdots & \notag \\
    a^n_{6}(x)p(x) + a^n_{6}(y)p(y) & + a^n_{6}(z)p(z)  \le b^n_6, \notag
\end{align} 
where $\boldsymbol{a}^n_i = [a^n_i(x),a^n_i(y),a^n_i(z)], ~i \in [1,..,6]$ is the outward unit normal vector on the $i_\text{th}$ face of the $n_\text{th}$ cuboid obstacle, and $b^n_i$ is a constant obtained from the dot product between $\boldsymbol{\alpha}^n_i$ and a known point on the plane which contains the $i_\text{th}$ face. This obstacle model has the flexibility to create 3D objects of varying dimensions, thus adequately representing real-world settings.

Suppose now that $\boldsymbol{p}$ describes the position of a target $\boldsymbol{x}^j_t$ at time-step $t$. This target, can avoid a potential collision with obstacle $\xi_n$ when the following condition holds:
\begin{equation}
    \exists~ i \in [1,..,6]: \text{dot}(\boldsymbol{a}^n_i, \boldsymbol{p}) > b^n_i,
\end{equation}
\noindent where $\text{dot}(a,b)$ is the dot product between vectors $a$ and $b$. In essence, we require that the target's position resides outside the convex-hull of obstacle $\xi_n, ~n \in [1,..,|\Xi|]$.

\section{Target Trajectory Planning}\label{sec:hypothesis_gen}

As we have already mentioned in Sec. \ref{ssec:target_dynamics}, for each target $j$ we consider the availability of the following information: a) its approximate initial location, i.e., we know that the state of target $j$ is initially distributed according to $\boldsymbol{x}^j_{0} \sim \mathcal{N}(\boldsymbol{\mu}^j_0,\Sigma^j_0)$, and b) its final destination, i.e., we know that target's $j$ objective is to move towards, and reach a specific goal region $\mathcal{G}^j$.
Based on these information, and in combination with a known map of the environment (i.e., in this work we use information regarding the position, and dimensions of various obstacles), the objective is to generate a hypothetical trajectory for each target, which can then be passively monitored through sensing, i.e., via the received bearing measurements, as discussed in Sec. \ref{ssec:agent_sensing_model}.

 To do that, the target trajectory hypothesis generation is formulated in this work as a model predictive control problem, where we seek to find target's $j$ hypothetical control inputs $U^j_t=\{\boldsymbol{u}^j_{t+\tau|t}\}, \forall \tau \in [0,..,T-1]$ inside a rolling finite planning horizon of length $T$ time-steps, which enable the guidance of the target to its goal region, subject to kinematic and collision avoidance constraints.

Let us denote the future hypothetical trajectory of target $j$ over a planning horizon of length $T$ time-steps, as $X^j_t = \{\boldsymbol{x}^j_{t+\tau+1|t}\}, \forall \tau \in [0,..,T-1]$, where the notation $\boldsymbol{x}_{t^\prime|t}$ is used here to denote the predicted target state at time-step $t^\prime$ which was generated at time-step $t$. Now, based on Eq. \eqref{eq:target_dynamics}, observe that the target trajectory $X^j_t$ is in fact a stochastic process, with each future target state $\boldsymbol{x}^j_{t+\tau+1|t}, \forall \tau$, to be distributed according to $\boldsymbol{x}^j_{t+\tau+1|t} \sim \mathcal{N}(\boldsymbol{\mu}^j_{t+\tau+1|t},\Sigma^j_{t+\tau+1|t})$, where $\boldsymbol{\mu}^j_{t+\tau+1|t}$, and $\Sigma^j_{t+\tau+1|t}$ are given by:
\begin{equation}
\begin{aligned}\label{eq:state_prediction}
    \!\!\!\!\!&\boldsymbol{\mu}^j_{t+\tau+1|t} = A^{\tau+1} \boldsymbol{\mu}^j_{t} + \sum_{k=0}^{\tau} A^{\tau-k} B \boldsymbol{u}^j_{t+k|t}, \\
    \!\!\!\!\!&\Sigma^j_{t+\tau+1|t} =  A^{\tau+1} \Sigma^j_{t} (A^\top)^{\tau+1} \!+\! \sum_{k=0}^{\tau} A^{\tau-k} Q (A^\top)^{\tau-k}.
 \end{aligned}
\end{equation}

\noindent Observe that Eq. \eqref{eq:state_prediction}, has been obtained from the recursive application of Eq. \eqref{eq:target_dynamics}. The parameters $\boldsymbol{\mu}^j_{t}$, and $\Sigma^j_{t}$ are respectively the mean, and covariance matrix of the target state at time-step $t$, which for time-step $t=0$ are given by $\boldsymbol{\mu}^j_0 $ and $\Sigma^j_0$ respectively.
In order to generate the trajectory which guides target $j$ to its goal region $\mathcal{G}^j$, the following cost function is minimized for the control inputs $U^j_t=\{\boldsymbol{u}^j_{t+\tau|t}\}, \forall \tau \in [0,..,T-1]$:
\begin{equation}\label{eq:mission_objective}
	\begin{aligned} 
    &\underset{U_t}{\arg\min} ~\mathbb{E}[\mathcal{J}^j(X^j_t,U^j_t)] = \|\boldsymbol{\mu}^{j,\text{pos}}_{t+T|t}-\mathcal{G}^j_o\|^2_2 \\ &
    ~~~~~~~~~~~~~~~~~~~~~~~+  \sum_{\tau=1}^{T-1} \|\boldsymbol{u}^j_{t+\tau|t}-\boldsymbol{u}^j_{t+\tau-1|t}\|^2_2, 
\end{aligned}
\end{equation}
\noindent where $\mathbb{E}$ is the expectation operator, $\|.\|_2$ is the 2-norm, $\boldsymbol{\mu}^{j,\text{pos}}_{t+T|t}$ is the predicted mean of the target's position at the end of the planning horizon computed with Eq. \eqref{eq:state_prediction}, and $\mathcal{G}^j_o$ is the centroid point of the goal region $\mathcal{G}^j$ on the ground. The second term in Eq. \eqref{eq:mission_objective} is used in order to minimize abrupt changes in the target's direction and speed, and thus produce more realistic smooth trajectories. The predicted target trajectory for agent $j$ is then generated with the guidance controller shown in Problem (P1). As shown in Problem (P1), at each time-step $t$ the optimal control inputs  $U^j_t=\{\boldsymbol{u}^j_{t+\tau|t}\}, \forall \tau \in [0,..,T-1]$ are computed over a rolling planning horizon of length $T$ time-steps, by solving an open-loop optimal control problem shown, which essentially drives the target to its goal region, while at the same time considering obstacle avoidance constraints. 

\begin{algorithm}
\begin{subequations}
\begin{align}
&\hspace*{-1mm}\textbf{Problem (P1)}: \texttt{Guidance Controller} &  \nonumber\\
& \hspace*{-1mm}~~~~~~~\underset{U^j_t}{\min} ~\mathbb{E}[\mathcal{J}^j(X^j_t,U^j_t)]&\hspace*{-3mm} \forall j\label{eq:objective_P1} \\
&\hspace*{-1mm}\textbf{subject to} ~ \tau \in [0,..,T-1] \textbf{:}  &\nonumber\\
&\hspace*{-1mm} \boldsymbol{\mu}^j_{t+\tau+1|t} = A^{\tau+1} \boldsymbol{\mu}^j_{t} + \sum_{k=0}^{\tau} A^{\tau-k} B \boldsymbol{u}^j_{t+k|t}, & \hspace*{-3mm} \forall \tau, j \label{eq:P1_1}\\
&\hspace*{-1mm} \boldsymbol{\mu}^j_{t} = \hat{\boldsymbol{\mu}}^j_{t|t-1}, \Sigma^j_{t} = \hat{\Sigma}^j_{t|t-1}, & \hspace*{-3mm} \forall j\label{eq:P1_2}\\
&\hspace*{-1mm} \text{dot}(\boldsymbol{a}^n_i, \boldsymbol{\mu}^{j,\text{pos}}_{t+\tau+1|t}) > b^n_{i} - H y^{j,n}_{\tau,i}, &\hspace*{-3mm} \forall \tau, j, n, i \label{eq:P1_3}\\
&\hspace*{-1mm} \sum_{i=1}^6 y^{j,n}_{\tau,i} \le 5, &\hspace*{-3mm} \forall \tau,j, n \label{eq:P1_4}\\
&\hspace*{-1mm} X^j_t \in \mathcal{X}, U^j_t\in \mathcal{U} &\hspace*{-3mm} \label{eq:P1_5}\\
&\hspace*{-1mm} y^{j,n}_{\tau, i} \in \{0,1\},~ n=[1,..,|\Xi|],~ i=[1,..,6] &\hspace*{-3mm} \notag
\end{align}
\end{subequations}
\end{algorithm}

Specifically, in Problem (P1) the constraints in Eq. \eqref{eq:P1_1}\eqref{eq:P1_2} compute the expected state of target $j$ (i.e., $\boldsymbol{\mu}^j_{t+\tau+1|t}$) inside the planning horizon, which has an associated covariance matrix $\Sigma^j_{t+\tau+1|t}$. Observe that the covariance matrix does not depend on the generated control inputs, and thus can be pre-computed as shown in Eq. \eqref{eq:state_prediction}. 
The constraints in Eq. \eqref{eq:P1_3}-\eqref{eq:P1_4} enable the generation of collision-free trajectories, by making sure that all targets avoid collisions with the obstacles in the environment. As a reminder, a collision with some obstacle $\xi_n,  n=[1,..,|\Xi|]$, which is represented as a cuboid, is avoided at time-step $t$ when the target state (i.e., its position coordinates) resides outside the convex hull of $\xi_n$ as explained in Sec. \ref{ssec:obstacle_model}. In order to enable this functionality we use the binary variable $y^{j,n}_{\tau,i} \in \{0,1\}$ which is activated i.e., $y^{j,n}_{\tau,i}=1$ when the inequality $\text{dot}(\boldsymbol{a}^n_i, \boldsymbol{\mu}^{j,\text{pos}}_{t+\tau+1|t}) > b^n_{i}$ is not satisfied for target $j$ with position $\boldsymbol{\mu}^{j,\text{pos}}_{t+\tau+1|t}$ at time-step $t+\tau+1|t$, and the $i_\text{th}$ face of the $n_\text{th}$ obstacle. In such cases the activation of $y^{j,n}_{\tau,i}$ makes the constraint shown in Eq. \eqref{eq:P1_3} valid with the utilization of a large positive constant $H \in \mathbb{Z}^+$. Now, as discussed in Sec. \ref{ssec:obstacle_model} a collision is avoided at time-step $t+\tau+1|t$ between the target $j$ with state $\boldsymbol{\mu}^{j,\text{pos}}_{t+\tau+1|t}$, and the obstacle $\xi_n$ when $\exists~ i \in [1,..,6]: \text{dot}(\boldsymbol{a}^n_i, \boldsymbol{\mu}^{j,\text{pos}}_{t+\tau+1|t}) > b^n_i$, which is achieved via the constraint in Eq. \eqref{eq:P1_4} by enforcing the binary variable $y^{j,n}_{\tau,i}$ to take the value of zero for at least one face i.e., $\exists i \in [1,..,6]: y^{j,n}_{\tau,i} = 0$. Finally, the constraints in Eq. \eqref{eq:P1_5} restrict the target's speed and control inputs within the desired limits. We should point out that Problem (P1) is a mixed integer quadratic program (MIQP), which can be solved efficiently using off-the-shelf optimization tools \cite{hvattum2012comparisons}.

\section{Autonomous UAV Control for Passive Multi-Target Monitoring} \label{sec:monitoring}

\subsection{Target State Estimation}

For each target $j$ the UAV agent maintains a Bayes filter \cite{sarkka2013bayesian}, which uses in order to compute, and recursively update over time its belief (i.e., a probability distribution) on the state of each target. This is shown in Eq. \eqref{eq:bayes_filter} where we denote as  $bel(\boldsymbol{x}^j_{t+1})$ the agent's initial belief on the state of target $j$ for the next time-step $t+1$, and with $\hat{bel}(\boldsymbol{x}^j_{t+1})$ we denote the posterior belief on the target's state after incorporating the received target measurements. 
\begin{subequations}\label{eq:bayes_filter}
\begin{align}
    bel(\boldsymbol{x}^j_{t+1})&= \int f(\boldsymbol{x}^j_{t+1}|\boldsymbol{x}^j_{t},\boldsymbol{u}^j_{t}) \hat{bel}(\boldsymbol{x}^j_{t}) d{\boldsymbol{x}^j_{t}} \label{eq:prediction} \\ 
    \hat{bel}(\boldsymbol{x}^j_{t+1}) &= \eta^{-1} g(\Phi^j_{t+1}|\boldsymbol{x}^j_{t+1},\boldsymbol{s}_{t+1}) bel(\boldsymbol{x}^j_{t+1})\label{eq:update}
\end{align} 
\end{subequations}

The agent's initial belief $bel(\boldsymbol{x}^j_{t+1})$ is computed through the prediction step shown in Eq. \eqref{eq:prediction}, where $f(\boldsymbol{x}^j_{t+1}|\boldsymbol{x}^j_{t},\boldsymbol{u}^j_{t})$ is the target state transition density which is governed by the target dynamics in Eq. \eqref{eq:target_dynamics}, and therefore is given by $f(\boldsymbol{x}^j_{t+1}|\boldsymbol{x}^j_{t},\boldsymbol{u}^j_{t}) = \mathcal{N}(A\boldsymbol{x}^j_{t} + B \boldsymbol{u}^j_{t},Q)$. On the other hand, $\hat{bel}(\boldsymbol{x}^j_{t})$ is the posterior belief of the current time-step i.e., $\hat{bel}(\boldsymbol{x}^j_{t}) = \mathcal{N}(\hat{\boldsymbol{\mu}}^j_{t},\hat{\Sigma}^j_{t})$, and thus $bel(\boldsymbol{x}^j_{t+1}) = \mathcal{N}(A\hat{\boldsymbol{\mu}}^j_{t} + B \boldsymbol{u}^j_{t},A\hat{\Sigma}^j_{t}A^\top +Q)$. Observe that this result is also obtained from Eq. \eqref{eq:state_prediction} by setting $\tau=0$, to obtain the one step look-ahead predictive density for the state of target $j$ computed at time-step $t$ i.e., $\boldsymbol{x}^j_{t+1|t} \sim \mathcal{N}(\boldsymbol{\mu}^j_{t+1|t},\Sigma^j_{t+1|t}) = bel(\boldsymbol{x}^j_{t+1})$.

Subsequently, at time-step $t+1$ the agent with state $\boldsymbol{s}_{t+1}$ receives from each target $j$ the measurement set $\Phi^j_{t+1}$, and updates its belief by computing $\hat{bel}(\boldsymbol{x}_{t+1})$ with the update step shown in Eq.\eqref{eq:update}. Specifically, $\eta = \int g(\Phi^j_{t+1}|\boldsymbol{x}^j_{t+1},\boldsymbol{s}_{t+1}) bel(\boldsymbol{x}^j_{t+1}) d\boldsymbol{x}^j_{t+1}$ is a normalizing constant, and the measurement likelihood function $g(\Phi^j_{t+1}|\boldsymbol{x}^j_{t+1},\boldsymbol{s}_{t+1})$ gives the likelihood that the agent with state $\boldsymbol{s}_{t+1}$ will receive at time-step $t+1$ the measurement set $\Phi^j_{t+1}$ from target $j$ with state $\boldsymbol{x}^j_{t+1}$.

To compute this likelihood function, first observe that the measurement set $\Phi^j_{t+1}$ contains a random number of random measurements i.e., multiple false-alarm measurements $\tilde{\phi}^j_{t+1,i} \in \Phi^j_{t+1}$ coming with a Poisson rate $\Lambda$, which are distributed according to $p_{\tilde{\phi}}(\tilde{\phi}^j_{t+1,i})$, and up to one target measurement $\hat{\phi}^j_{t+1} \in \Phi^j_{t+1}$ which is received with probability $p_D$, and which is distributed according to $c(\hat{\phi}^j_{t+1}) = \mathcal{N}(\hat{\phi}^j_{t+1};\ell(\boldsymbol{x}^j_{t+1},\boldsymbol{s}_{t+1}),\sigma^2_{\phi})$ as discussed in Sec. \ref{ssec:agent_sensing_model}. That said, the measurement likelihood function is derived as:
\begin{align} \label{eq:rfs_like}
    &g(\Phi^j_{t+1}|\boldsymbol{x}^j_{t+1},\boldsymbol{s}_{t+1})=(1-p_D) n^j! \Psi(n^j;\Lambda) \prod_{\phi \in \Phi_{t+1}} p_{\tilde{\phi}}(\phi) \notag \\
   & + (n^j-1)! \Psi(n^j-1;\Lambda) p_D \underset{\phi \in \Phi_{t+1}}{\sum} c(\phi)  \underset{ \begin{subarray}{l}
  \varphi \in \Phi_{t+1} \\
  \varphi \ne \phi
  \end{subarray}}{\prod} p_{\tilde{\phi}}(\varphi)
\end{align}
\noindent where $n^j = |\Phi^j_{t+1}|$ is the total number of received measurements, and $\Psi(n^j;\Lambda)$ is probability mass function of the Poisson distribution with rate parameter $\Lambda$, and input argument $n^j$. Therefore, the first term in Eq. \eqref{eq:rfs_like} computes the event of receiving at time-step $t+1$ exactly $n^j$ false-alarm measurements (i.e., $\Psi(n^j;\Lambda) \prod_{\phi \in \Phi_{t+1}} p_{\tilde{\phi}}(\phi)$), and no measurement from target $j$, i.e., the target is not detected with probability $(1-p_D)$; and the factor $n^j!$ accounts for all possible permutations of the measurements in the set. On the other hand, the second term in Eq. \eqref{eq:rfs_like} accounts for the event where the measurement set $\Phi^j_{t+1}$ contains a single target measurement $\hat{\phi}$ with likelihood $p_D c(\hat{\phi})$, and $(n-1)$ false-alarm measurements.
Finally, the posterior mean $(\hat{\boldsymbol{\mu}}^j_{t+1})$ and covariance $(\hat{\Sigma}^j_{t+1})$ of the state of target $j$ for time-step $t+1$ is extracted from $\hat{bel}(\boldsymbol{x}^j_{t+1})$, which are used to initialize the guidance controller for the next time-step, and subsequently, the recursion shown in Eq. \eqref{eq:bayes_filter} is repeated for the next time-step.



\subsection{Monitoring Control}
\begin{figure*}
	\centering
	\includegraphics[width=\textwidth]{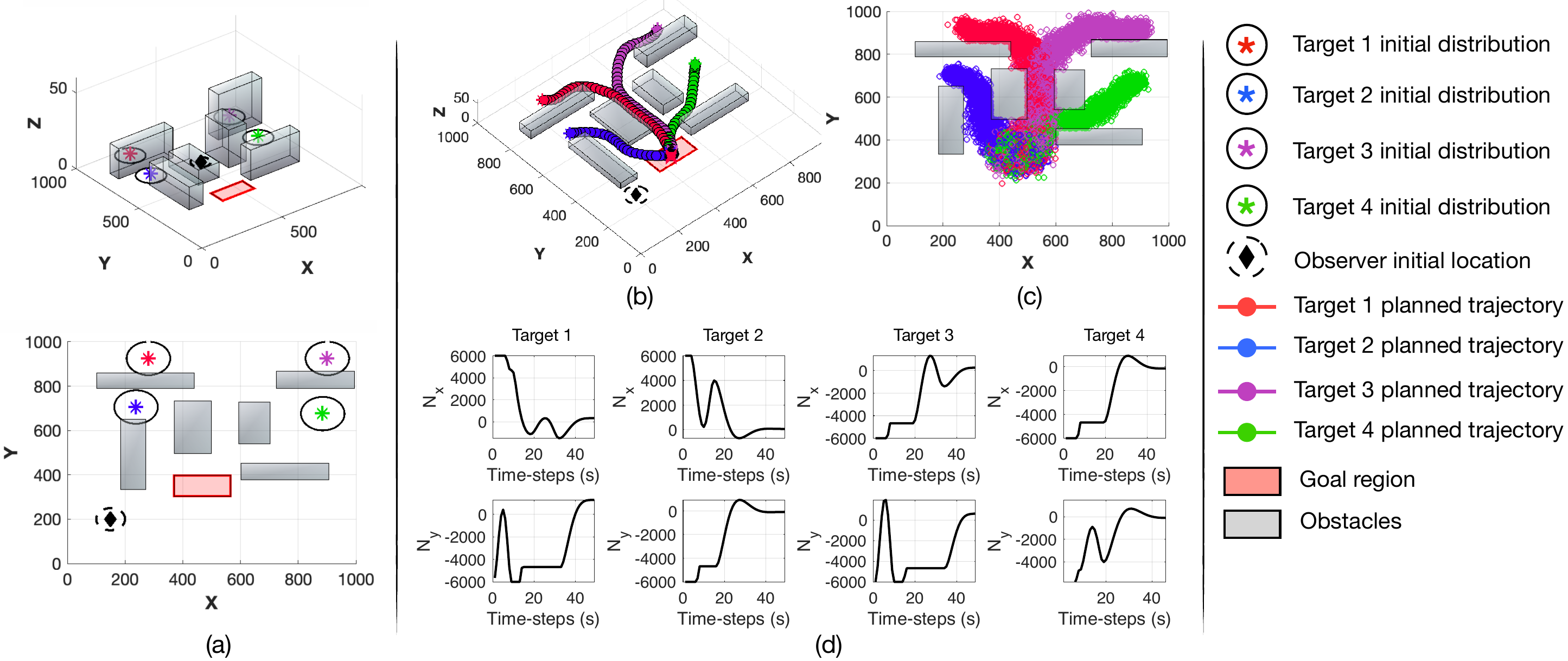}
	\caption{The figure illustrates the proposed target trajectory hypothesis generation approach, which is realized with the guidance controller shown in Problem (P1), and allows the 4 targets to be guided to the goal region while avoiding collisions with the obstacles in the environment.}
	\label{fig:res1}
	\vspace{-4mm}
\end{figure*}
In order to optimize the monitoring performance at time-step $t+1$ for a particular target $j$  it suffices to select the agent's next state $\hat{\boldsymbol{s}}_{t+1} \in \mathcal{S}_{t+1}$, which will result in the future measurement set $\Phi^j_{t+1}$, which maximizes the observability of the target state. This strategy however cannot be applied directly since the measurement set $\Phi^j_{t+1}$ becomes available only after the agent moves to its new state $\hat{\boldsymbol{s}}_{t+1}$. To overcome this limitation, we follow the procedure described next: For each admissible agent state $\boldsymbol{s}_{t+1} \in \mathcal{S}_{t+1}$, we generate for each target $j$ the hypothetical ideal (i.e., noise-free, no false-alarms) measurement set $Z^j_{t+1} = \{z^j_{t+1}\}$, which would have been received if the agent moves at time-step $t+1$ to state $\boldsymbol{s}_{t+1}$, and target $j$ is distributed according to $ bel(\boldsymbol{x}^j_{t+1})$ (computed with Eq. \eqref{eq:prediction}), with expected position denoted as $\mu^{j,\text{pos}}_{t+1}$. That said, the hypothetical measurement set $Z^j_{t+1} = \{z^j_{t+1}\}$ is generated as:
\begin{equation} \label{eq:measurement_model2}
	z^j_{t+1} = \tan^{-1}\left(\frac{\mu^{j,\text{pos}}_{t+1}(x)-s_{t+1}(x)}{\mu^{j,\text{pos}}_{t+1}(y) - s_{t+1}(y)}\right).
\end{equation}

\noindent Then, for each pair $(\boldsymbol{s}_{t+1},z^j_{t+1})_i, ~i \in [1,..,|\mathcal{S}_{t+1}|]$ we compute the pseudo-posterior distribution $\tilde{bel}(\boldsymbol{x}^j_{t+1},\boldsymbol{s}_{t+1},z^j_{t+1})_i$ according to Eq. \eqref{eq:update}, where the measurement likelihood function $g(z^j_{t+1})$ is now given by $g(z^j_{t+1}|\boldsymbol{x}^j_{t+1},\boldsymbol{s}_{t+1}) = \mathcal{N}(z^j_{t+1};\ell(\boldsymbol{x}^j_{t+1},\boldsymbol{s}_{t+1}),\sigma^2_{\phi})$. Finally, we extract the pseudo-posterior target state mean and covariance $(\tilde{\boldsymbol{\mu}}^j_{t+1},\tilde{\Sigma}^j_{t+1})_i$. The optimal state $\hat{\boldsymbol{s}}_{t+1}$ of the UAV agent for time-step $t+1$ which achieves optimized monitoring performance is then obtained as:
\begin{equation} \label{eq:optimization}
	\hat{\boldsymbol{s}}_{t+1} = \underset{\boldsymbol{{s}} \in \mathcal{S}_{t+1}}{\arg\min} \sum_{j=1}^{M} \tr\left( \tilde{\Sigma}^j_{t+1}(\boldsymbol{{s}}) \right),
\end{equation}
\noindent where $\tr(\Sigma)$ is the trace of matrix $\Sigma$, and the notation $\tilde{\Sigma}^j_{t+1}(\boldsymbol{{s}})$ denotes the covariance matrix associated with the pseudo-posterior distribution of the state of the $j_\text{th}$ target, which was obtained under the assumption that the agent moved to state $\boldsymbol{s}$ at time-step $t+1$. Once the the optimization problem of Eq. \eqref{eq:optimization} is solved, the agent moves to its new state $\hat{\boldsymbol{s}}_{t+1}$, where the actual target measurements $\Phi^j_{t+1}, \forall j$ are received, and subsequently the posterior distribution on the target states is computed with Eq. \eqref{eq:update} as explained earlier. 


\section{Evaluation} \label{sec:Evaluation}

\subsection{Simulation Setup} \label{ssec:sim_setup}

To evaluate the proposed approach we have used the following simulation setup. The surveillance area $\mathcal{E} \subset \mathbb{R}^3$ is given by a cube with a total volume of 1$\text{km}^3$. The target dynamics are given by Eq. \eqref{eq:target_dynamics} with $\Delta t=1$s, $\varepsilon=0.2$, and $m=1300$kg, and are the same for all $M=4$ targets. The process noise $\boldsymbol{\nu}_t$ is distributed according to $\boldsymbol{\nu}_t \sim \mathcal{N}(0,Q)$, with $Q = \text{diag}([30~30~eps~3~3~eps])$, where $eps$ is a very small number i.e., $eps = 1\text{E}-10$, which indicates our knowledge that the targets evolve on the ground plane. Initially it is assumed that the four targets are distributed according to $\boldsymbol{x}^1_0 \sim \mathcal{N}(\boldsymbol{\mu}^1_0,\Sigma_0)$, $\boldsymbol{x}^2_0 \sim \mathcal{N}(\boldsymbol{\mu}^2_0,\Sigma_0)$, $\boldsymbol{x}^3_0 \sim \mathcal{N}(\boldsymbol{\mu}^3_0,\Sigma_0)$, and $\boldsymbol{x}^4_0 \sim \mathcal{N}(\boldsymbol{\mu}^4_0,\Sigma_0)$, where $\boldsymbol{\mu}^1_0 = [281, 925, 0]$m, $\boldsymbol{\mu}^2_0 = [238, 706, 0]$m, $\boldsymbol{\mu}^3_0 = [901, 925, 0]$m, and $\boldsymbol{\mu}^4_0 = [885, 676, 0]$m. The covariance matrix $\Sigma_0$ is given by $\Sigma_0=\text{diag}([200~200~eps~20~20~eps])$ for all targets. 

The control input $\boldsymbol{u}_t$ is bounded in the $x$, and $y$ dimensions inside the interval $[-6000,6000]$N for all targets, and in the $z$ dimension is zero. The targets can reach a ground speed of up to 16$\text{m}/\text{s}$. The agent dynamics are given by Eq. \eqref{eq:controlVectors} with $\Delta_r=5$m, $N_\theta = 15$, $N_r=4$, and $h=40$m. In addition, the measurement noise $w_t$ is distributed according to $w_t \sim \mathcal{N}(0,\sigma^2_\phi)$, with $\sigma_\phi=1\deg$, and the target detection probability is set to $p_D=0.95$. The false-alarms are uniformly distributed inside the measurement space $(-\pi,\pi]$, and arrive with a Poisson rate $\Lambda=1$. Finally, we note that the stochastic filtering recursion in Eq. \eqref{eq:bayes_filter} has been implemented as a particle filter \cite{gustafsson2010particle} mainly for handling the non-linear measurement model i.e., Eq. \eqref{eq:measurement_model}, and the guidance problem i.e., Problem (P1), was solved with the Gurobi's MIQP solver.

\subsection{Performance Evaluation}
 Figure \ref{fig:res1}(a) shows in 3D and top-down views, the initial position of targets $\boldsymbol{x}^1$, $\boldsymbol{x}^2$, $\boldsymbol{x}^3$, and $\boldsymbol{x}^4$ which are marked with a red, blue, purple and green $\star$ respectively. The initial covariance of the target states is drawn as an error-ellipse around the mean of the target location as shown. The obstacles in the environment are shown as grey coloured cuboids, and the goal region, which in this scenario is the same for all targets, is shown with the red rectangular region. The UAV agent is initialized in this example at $\boldsymbol{s}_0 = [150, 200, 40]$m, as shown with the black $\diamond$. Figure~\ref{fig:res1}(b) shows the output of the proposed guidance controller as depicted in Problem (P1), which has generated the hypothetical planned trajectories for the 4 targets over a planning horizon $T=50$ time-steps. As shown in the figure, this optimization allows the targets to avoid the obstacles in the environment, and based on their mobility capabilities to reach the goal region as soon as possible. Fig.~\ref{fig:res1}(c) shows the uncertainty on the targets' states over the planning horizon as computed at time-step $t=1$, with the Eq. \eqref{eq:state_prediction}. In particular, the figure shows the target position as particles sampled from $\mathcal{N}(\boldsymbol{\mu}^j_{1+\tau+1|1},\Sigma^j_{1+\tau+1|1}), \forall \tau, \forall j$. Finally, Fig. \ref{fig:res1}(d), shows the optimal control inputs ($x$ and $y$ dimensions) over the planning horizon that guide the targets to the goal region, while producing smooth trajectories without abrupt changes in the speed and direction. 

\begin{figure}
	\centering
	\includegraphics[width=\columnwidth]{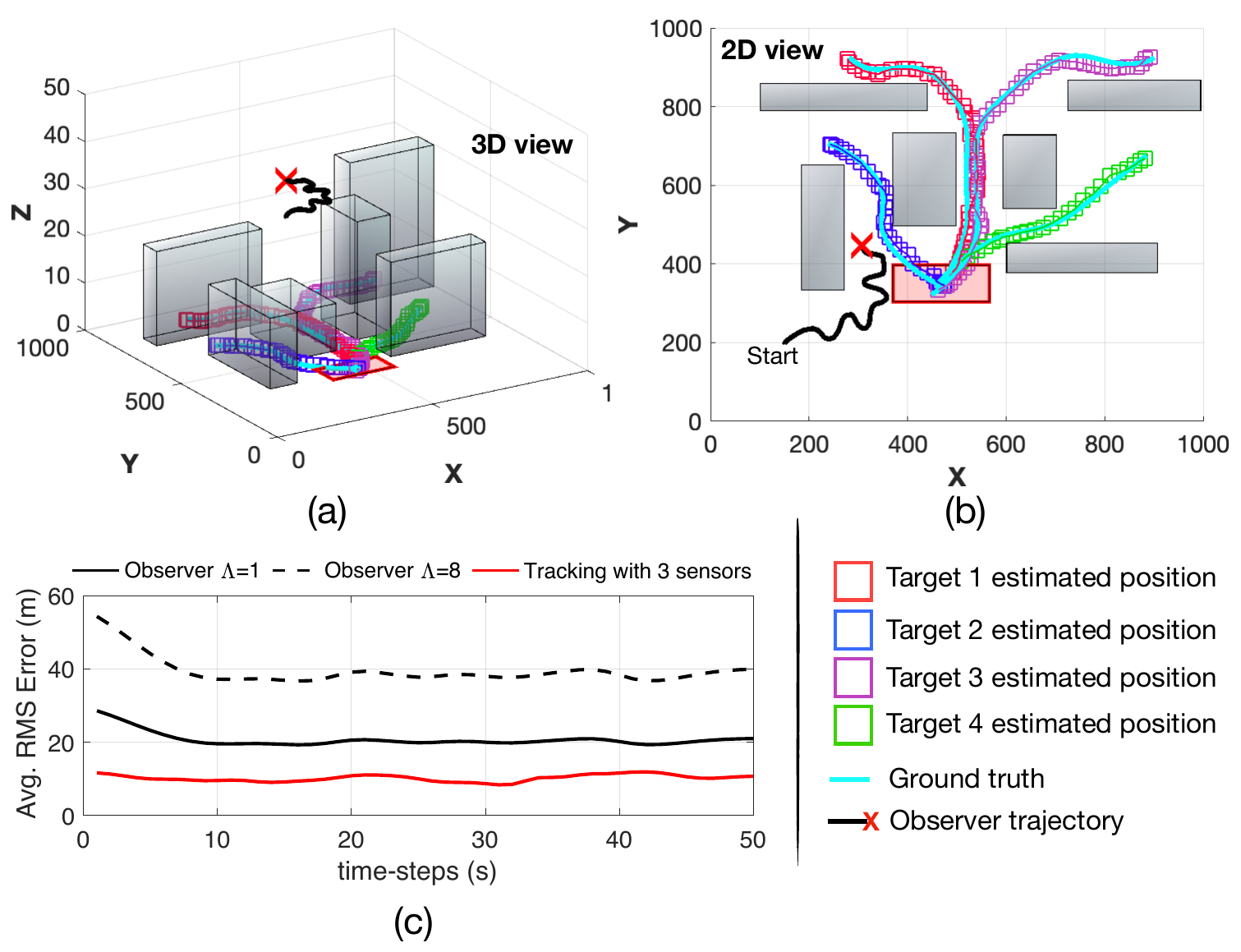}
	\caption{(a)(b) The figure shows the optimized trajectory of the UAV agent which maximizes the target monitoring performance, (c) the average positional RMSE obtained during 100 Monte-Carlo trials.}
	\label{fig:res2}
	\vspace{-3mm}
\end{figure}

Next we demonstrate the performance of the proposed approach for the task of passively monitoring the four ground targets. The objective now becomes the selection of the optimal UAV control inputs at each time-step such that the collective uncertainty on the target states is minimized. To achieve this, the UAV agent makes a prediction on the target next states as discussed in Sec. \ref{sec:hypothesis_gen}, and then uses the received target measurements to update those predictions using the filtering procedure discussed in Sec. \ref{sec:monitoring}. Figure \ref{fig:res2}(a)(b) show the result of the optimization problem in Eq. \eqref{eq:optimization} i.e., the UAV's optimal trajectory which maximizes the monitoring performance i.e., minimizes the uncertainty on the target states. Essentially, the UAV agent seeks at each time-step to select its next state from which will obtain the most informative bearing measurement, and which in turn will allow the estimation of the target state. We define the root mean square error (RMSE) on the target position at time-step $t$ as $\epsilon_t = \sqrt{N^{-1}\sum_{n=1}^N||\hat{\boldsymbol{x}}^\text{pos}_t(n) - \boldsymbol{x}^\text{pos}_t||_2^2}$, where $||.||_2^2$ is the squared 2-norm, $N$ is the number of Monte-Carlo trials, $\hat{\boldsymbol{x}}^\text{pos}_t(n)$ denotes the estimated $(x,y)$ target coordinates at time-step $t$ on the $n_\text{th}$ trial, and $\boldsymbol{x}^\text{pos}_t$ is the true target position at the same time-step. Figure \ref{fig:res2}(c) shows the average positional RMSE obtained for tracking the four targets during the scenario depicted in Fig. \ref{fig:res2}(a)(b). This scenario was simulated for $N=100$ trials, where in each trial the UAV initial position was randomly initialized inside the surveillance area. This result is then compared with the positional error obtained from a 3-sensor tracking system. Specifically, we assume that three fixed direction-finding sensors located at $[150,200]$m, $[800,200]$m, and $[500,900]$m, receive three bearing measurements from each target at each time-step, and localize the targets according to the procedure discussed in Sec. \ref{sec:monitoring} by combining their individual measurement likelihood functions. The measurement noise profile in this case is as discussed in Sec. \ref{ssec:sim_setup}, however without false-alarms. As shown in the graph, although the 3-sensor system achieves better results (note that in this case the target state is fully observable), the proposed single sensor system by optimizing the measurement collection process, achieves comparable performance (i.e., solid black line) despite the presence of false-alarms. Finally, the black dotted-line shows what is the achievable performance of the proposed approach in scenarios with higher false-alarm rates i.e., $\Lambda=8$. Although, the rate of false-alarms degrades the overall monitoring performance as shown in the figure, the targets can still be tracked with a reasonable accuracy, which can be adequate for certain application domains.

\section{Conclusion} \label{sec:conclusion}
In this work we propose a joint estimation and control approach for passively monitoring multiple targets of interest in challenging conditions (i.e., environments with obstacles, and false-alarm measurements) with a single UAV agent equipped with a direction-finding sensor. Model predictive control is used for generating hypothetical target trajectories inside a rolling finite planning horizon, which are then refined through stochastic filtering. In particular, we show how the agent's path can be optimized in order to minimize the collective uncertainty over the target states. Future work includes the implementation of the proposed approach on UAV hardware platforms and its validation in real-world settings.



\bibliographystyle{IEEEtran}
\bibliography{main} 

\end{document}